\documentclass{article}

\usepackage{amsmath,amsfonts,braket,graphicx}

\newcommand{\ketbra}[1]{\ket{#1}\bra{#1}}

\title{Security of a Semi-Quantum Protocol Where Reflections Contribute to the Secret Key}
\author{Walter O. Krawec\\\small{Iona College}\\\small{New Rochelle, NY 10801 USA}\\\small{\texttt{walter.krawec@gmail.com}}}

\begin{document}
\maketitle

\begin{abstract}
In this paper we provide a proof of unconditional security for a semi-quantum key distribution protocol introduced in a previous work.  This particular protocol demonstrated the possibility of using $X$ basis states to contribute to the raw key of the two users (as opposed to using only direct measurement results) even though a semi-quantum participant cannot directly manipulate such states.  In this work we provide a complete proof of security by deriving a lower bound of the protocol's key rate in the asymptotic scenario.  Using this bound we are able to find an error threshold value such that for all error rates less than this threshold, it is guaranteed that $A$ and $B$ may distill a secure secret key; for error rates larger than this threshold, $A$ and $B$ should abort.  We demonstrate that this error threshold compares favorably to several fully quantum protocols.  We also comment on some interesting observations about the behavior of this protocol under certain noise scenarios.
\end{abstract}
\section{Introduction}
Quantum Key Distribution (QKD) protocols allow two users: Alice ($A$) and Bob ($B$) to establish a secret key in the presence of an all powerful adversary.  Since the original BB84 protocol \cite{QKD-BB84}, several protocols have been developed including B92 \cite{QKD-B92}, SARG04 \cite{QKD-SARG04}, three state BB84 \cite{QKD-BB84-three-state}, and many others (see \cite{QKD-survey} for a general survey).  These protocols, however, assume that $A$ and $B$ are both able to perform certain quantum operations (e.g., prepare and measure qubits in a variety of bases).

Semi-Quantum Key Distribution (SQKD) protocols, first introduced in 2007 \cite{SQKD-first}, attempt to achieve the same end (establishment of a secret key secure against an all powerful adversary), when one of the two users (typically $B$) is limited or ``classical'' in nature (what is meant by this shall be discussed momentarily).  Since their creation, several SQKD protocols have been proposed \cite{SQKD-second,SQKD-3,SQKD-4,SQKD-lessthan4,SQKD-multi1,SQKD-Single-Security,SQKD-MultiUser}.

A SQKD protocol typically operates by having the quantum user $A$ send a qubit prepared in an arbitrary basis.  The qubit travels to the classical user $B$ who is limited to performing one of two operations:
\begin{enumerate}
  \item He may \emph{measure and resend} the qubit; measuring only in the computational $Z = \{\ket{0}, \ket{1}\}$ basis and resending his result to $A$.  That is, if he measures $\ket{r}$ ($r \in \{0,1\}$), then he will send a qubit $\ket{r}$ to $A$.
  \item He may \emph{reflect} the qubit; the qubit then passes through $B$'s lab undisturbed and returns to $A$.  $B$ learns nothing of its state in this case.
\end{enumerate}
Regardless of $B$'s choice, a qubit returns to $A$ who is then free to perform any quantum operation on it (e.g., measure in an arbitrary basis).

Due to the reliance on a two-way quantum communication channel (one which permits a qubit to travel from $A$ to $B$ and then return from $B$ to $A$), the security analysis of SQKD protocols has been limited due to the fact that the attacker Eve ($E$) is now allowed two opportunities to attack the qubit, thus greatly increasing the complexity of the security analysis.  For this reason, most security proofs for SQKD protocols have been limited to the notion of \emph{robustness}.  This concept, introduced in \cite{SQKD-first} defines a protocol as robust if for any attack which allows $E$ to gain information on $A$ or $B$'s key with non-zero probability, must necessarily induce a disturbance which may be detected by either $A$ or $B$ with non-zero probability.  Nothing, however, is said about the relationship between the noise in $E$'s attack and the information gained.

Recently, however, the state of this situation has been improving.  In \cite{SQKD-information,SQKD-cl-A}, a relationship was derived between the probability of disturbance and the amount of information gained by $E$ assuming the latter is limited to performing \emph{individual attacks} (attacks where $E$ will perform the same operation each iteration of the protocol and will measure her ancilla before $A$ and $B$ use their key for any purpose).  Recently, we have managed to prove the unconditional security (making no assumptions on the type of attack employed by $E$) of several SQKD protocols by devising several different proof techniques for handling the complexity caused by the two-way quantum channel \cite{SQKD-MultiUser,SQKD-Krawec-SecurityProof}.  These techniques are also used to devise lower bounds on the key rate expression in the asymptotic scenario (to be defined shortly) for several protocols, namely Boyer et al.'s original protocol \cite{SQKD-first}, the single-state SQKD protocol introduced in \cite{SQKD-lessthan4}, and the mediated SQKD protocol introduced in \cite{SQKD-MultiUser}.

In \cite{SQKD-Single-Security}, we introduced a new semi-quantum key distribution protocol which was the first to permit reflections (and thus $X$ basis states) to contribute towards the raw key (note that classical $B$ cannot directly manipulate $X$-basis states).  However, in that paper we proved only its robustness - that is, we showed if an attacker gained information on the raw key, she could be detected with non-zero probability.  While at the time robustness was the primary definition of security for semi-quantum protocols, lately, however, this has not been the case.  In \cite{SQKD-Krawec-dissertation}, we provided a proof of its unconditional security by comparing it to the B92 \cite{QKD-B92} protocol; however, that proof technique led to very pessimistic noise tolerance levels along with an extreme sensitivity to the noise in the forward channel.  In this paper we revisit our protocol and use the technique developed in \cite{SQKD-Krawec-SecurityProof} (where we proved the security of Boyer et al.'s original SQKD protocol) to prove its security.  To do so, we will also utilize the notion of restricted collective attacks developed in \cite{SQKD-Single-Security} along with a suitable adaptation of the proof mechanism used in \cite{SQKD-Krawec-SecurityProof}.  This new proof provides a far more optimistic bound on its key rate and noise tolerance levels.

Our proof in this paper involves only the perfect qubit scenario.  We do not consider such attacks as multi-photon attacks for instance.  This is reasonable: indeed, security proofs for fully quantum protocols began by considering perfect qubit sources and only later began to analyze implementation issues.  In the future, analyzing more practical implementations, and security against the attacks (and potential remedies) mentioned in \cite{SQKD-photon-tag,SQKD-photon-tag-comment} will become more important.  However, we feel that the work presented in this paper and in others, in developing the techniques to prove these protocols secure, will be useful, not only in this semi-quantum setting, but also in analyzing other quantum protocols, semi or otherwise, requiring the use of a two-way quantum communication channel.  The analytical techniques developed and applied here may find broader application beyond the realm of semi-quantum and towards fully quantum protocols relying on a two-way channel.  They may also be helpful when considering security beyond the perfect qubit scenario.

\subsection{The Protocol}
The protocol we consider is exactly the one described in \cite{SQKD-Single-Security} with only one small modification mentioned later.  This protocol, besides being semi-quantum (i.e, one of the users - $B$ in this case - is limited to measuring and resending in the $Z$ basis or reflecting qubits) is also a single-state protocol.  These protocols, first introduced in \cite{SQKD-lessthan4}, place a further restriction on the quantum user $A$: namely, $A$ must send a single, publicly known, qubit state each iteration.

The quantum communication stage of the protocol is as follows:
\begin{enumerate}
  \item $A$ sends the state $\ket{+} = \frac{1}{\sqrt{2}}(\ket{0}+\ket{1})$.
  \item $B$ will choose a random value $k_B \in \{0,1\}$ to be his candidate raw key bit for this iteration.
  \begin{itemize}
    \item If $k_B = 0$ (with probability $1/2$), $B$ will reflect the qubit.
    \item If $k_B = 1$ he will measure and resend the qubit (i.e., he will measure in the $Z$ basis, receive outcome $\ket{r}$ for $r \in \{0,1\}$ and send the qubit $\ket{r}$ to $A$).  He saves his measurement result as $m_B$ (i.e., $m_B = r$).
  \end{itemize}
  
  \item $A$ now chooses to measure in the $Z$ or $X$ basis (the $X$ basis consists of those states $\ket{\pm} = \frac{1}{\sqrt{2}}(\ket{0} \pm \ket{1})$).
  \begin{itemize}
    \item If she chose the $X$ basis (with probability $1/2$), and her measurement result is $\ket{-}$, she sets her raw key bit to be $k_A = 1$
    \item If she chose the $Z$ basis and her measurement result is $\ket{1}$ she sets $k_A = 0$.
    \item Otherwise she sets $k_A = -1$.
  \end{itemize}
\end{enumerate}

After running the above process $N$ times, for $N$ sufficiently large, $A$ and $B$ will use the public authenticated channel to discard certain iterations.  $B$ will tell $A$ to discard all iterations where he measured $\ket{1}$.  $B$ will next choose a suitable proportion of randomly chosen iterations where $k_B = 0$ (i.e., those iterations he reflected) and discard them so that the probability that $k_B = 0$ is equal to the probability that $k_B = 1$.  This is done to ``balance'' his key (in the absence of any noise, he will discard half of the iterations he reflected - this is due to the fact that he discarded half of the iterations he measured due to him measuring $\ket{1}$; however $E$'s attack may ``bias'' his measurement results).  It is the only step we added from our original protocol in \cite{SQKD-Single-Security} - it is not necessary, but it does simplify the algebra in our proof.  Note that our analysis could still be carried out without this step - it is purely for algebraic simplification purposes.  Finally, $A$ will tell $B$ to discard all iterations where $k_A = -1$ (i.e., all iterations where she measured $\ket{0}$ or $\ket{+}$).

Note it is easy to see the protocol is correct.  That is, in the absence of noise, conditioning on the event that $B$ does not discard the iteration, the qubit leaving his lab is $\ket{+}$ (if $k_B = 0$) or $\ket{0}$ (if $k_B = 1$).  Thus, if $A$ measures $\ket{-}$ it must be that $B$ set $k_B = 1$ (if he reflects, $A$ should always measure $\ket{+}$); otherwise if she measures $\ket{1}$ it must be that $B$ reflected (i.e., $k_B = 0$).  The observant reader will note the similarities between this protocol and the B92 \cite{QKD-B92} protocol - a similarity first commented on in \cite{SQKD-Krawec-dissertation}.  Indeed, while Boyer et al.'s original SQKD protocol in \cite{SQKD-first} is often considered the semi-quantum version of BB84, our protocol may be considered the semi-quantum version of B92.

\section{Security Proof}
We will first assume $E$ is limited to performing \emph{collective attacks}.  These are attacks where $E$ performs the same operation each iteration of the protocol but is free to postpone the measurement of her ancilla until any future time of her choosing (this is in contrast to \emph{individual attacks} where she is forced to measure her ancilla immediately).  After proving security in this case, we will show security against \emph{general attacks} - attacks where there are no restrictions placed on $E$ other than those imposed by the laws of physics.

Following the completion of $N$ ``successful'' iterations of the protocol (i.e., those iterations that are not discarded) and, assuming collective attacks, the state of the joint quantum system $\rho_{ABE}$ (a density operator acting on $A$, $B$, and $E$'s Hilbert spaces - all of which, without loss of generality, are assumed to be finite dimensional) is of the form $\rho_{ABE} = \sigma_{ABE}^{\otimes N}$, where $\sigma_{ABE}$ models the joint system after one iteration.  Following the quantum communication stage, $A$ and $B$ will run an error correcting protocol and a privacy amplification protocol (see \cite{QKD-survey} for these standard processes) which will result in a secret key of size $\ell(N) \le N$ (possibly $\ell(N) = 0$ if $E$ has too much information).  Given this, it was shown in \cite{QKD-renner-keyrate,QKD-renner-keyrate2,QKD-Winter-Keyrate} that the key rate in the asymptotic scenario, denoted $r$, is:
\begin{equation}\label{eq:keyrate}
r = \lim_{N\rightarrow \infty}\frac{\ell(N)}{N} = \inf(S(B|E) - H(B|A)),
\end{equation}
where $S(B|E)$ is the conditional von Neumann entropy of $B$'s system conditioned on $E$ (defined $S(B|E) = S(BE) - S(E)$ where $S(\cdot)$ is the von Neumann entropy) while $H(B|A)$ is the classical conditional entropy of $B$'s system conditioned on $A$.  The infimum is over all collective attacks which induce the observed statistics (e.g., which induce the observed error rate).  This expression is rather intuitive: it states that the key rate is the difference between $E$'s uncertainty on $B$'s raw key (which should be high) and $A$'s uncertainty of $B$'s key (which should be low).  The infimum is required since there are infinitely many attacks which induce a certain error rate and we must assume that $E$ chooses the one which increases her information.  Note that we are using reverse reconciliation \cite{QKD-survey} here which, as commented in \cite{SQKD-Krawec-SecurityProof}, seems the more natural choice for these two-way semi-quantum protocols.  An open question remains to bound the key rate using direct reconciliation (there we have $S(A|E) - H(A|B)$).

In this paper, we will find a lower bound on this value $r$.  Our bound will be a function of certain parameters that may be estimated by $A$ and $B$.

\subsection{Modeling the Protocol}
To compute a bound on the key rate, we must first describe the quantum system after one successful iteration of the protocol.  Successful, here, meaning that $A$ and $B$ use this iteration to contribute towards their raw key (i.e., neither $A$ nor $B$ discard the iteration).  In particular, one of the following events occur:
\begin{enumerate}
  \item $B$ reflects (and does not later discard - recall he will choose a suitable portion of reflections and discard them so as to balance his raw key prior to $A$'s acceptance) and $A$ measures $\ket{1}$.
  \item $B$ measures and resends $\ket{0}$ and $A$ measures $\ket{-}$.
\end{enumerate}

In \cite{SQKD-Single-Security}, we showed that, for any single-state semi-quantum protocol (i.e., a protocol where $A$ sends the same qubit state each iteration and this state is public knowledge), and any collective attack $(U_F,U_R)$ (where $U_F$ is the unitary attack operator used by $E$ in the forward direction, while $U_R$ is the operator used in the reverse channel), there exists an equivalent restricted collective attack of the form $(b,U)$ where $b \in [-1/2, 1/2])$ and $U$ is a unitary operator acting on the qubit and $E$'s private quantum memory.  This attack works as follows:
\begin{enumerate}
  \item First, $E$ will capture the qubit sent from $A$.  Since it is always the same state, we have $E$ discard the qubit and prepare one of her own in the form:
  \[
  \ket{e} = \sqrt{\frac{1}{2}+b}\ket{0} + \sqrt{\frac{1}{2}-b}\ket{1}.
  \]
  She then sends this qubit $\ket{e}$ to $B$ (observe it is not entangled with her private quantum memory at this point).
  \item After $B$'s operation, the qubit returns to $A$; $E$ will first capture it, however, and probe it using unitary operator $U$ acting on the qubit and entangling it with her private quantum memory.  The qubit is then forwarded to $A$.
\end{enumerate}

It was proven that the resulting density operator under attack $(U_F, U_R)$ is equal to the density operator if attack $(b,U)$ is employed; thus, as far as $A$, $B$, or $E$ is concerned the two attacks are equivalent.  Furthermore, the resulting key rate expression is exactly the same in either case.  Thus, to prove security against collective attacks, it suffices to prove security agains these restricted attacks.  Later, we will prove security against general attacks.

Fix a restricted collective attack $(b, U)$.  We will assume that $b \in (-1/2, 1/2)$ (i.e. it is not equal to the extremes of $\pm1/2$).  As we will see later, $b$ contributes to the error rate of the protocol and if $b$ is too far from zero, the noise level is too high anyway so $A$ and $B$ should abort (note that $b$ is a parameter that $A$ and $B$ may estimate).

Due to the fact that $B$ will reject a suitable number of iterations when he reflects so as to balance the probability that, prior to $A$'s acceptance his raw key is $0$ or $1$, we may write the state of the system, following $B$'s operation, and conditioning on his acceptance, as follows:
\[
\rho = \frac{1}{2}\ketbra{0}_B \otimes \ketbra{e}_T + \frac{1}{2}\ketbra{1}_B \otimes \ketbra{0}_T,
\]
where $\ket{e} + \sqrt{1/2+b}\ket{0} + \sqrt{1/2-b}\ket{1}$, is the qubit sent from $E$ in the forward channel.  Of course $B$ can only know how many reflection iterations to discard after estimating $b$; thus he rejects these iterations after the quantum communication stage of the protocol.  However, whether we condition on this event now or later makes no difference to the resulting density operator.  Conditioning on it now, however, does simplify the following algebra.

The transit qubit then returns, from $B$, to $A$, but it is first intercepted by $E$ who will probe the qubit using unitary operator $U$.  This operator, which acts on $\mathcal{H}_T \otimes \mathcal{H}_E$, acts on basis states as follows:
\begin{align*}
U\ket{0} &= \ket{0,e_0} + \ket{1,e_1}\\
U\ket{1} &= \ket{0,e_2} + \ket{1,e_3}\\
U\ket{+} &= \ket{+,f_0} + \ket{-,f_1}\\
U\ket{-} &= \ket{+,f_2} + \ket{-,f_3}.
\end{align*}
Here, the $\ket{e_i}$ are arbitrary, not necessarily normalized, nor orthogonal, states in $\mathcal{H}_E$.  The $\ket{f_i}$ are states in $\mathcal{H}_E$ which depend linearly on the $\ket{e_i}$ states.  In particular, we have:
\begin{align}
\ket{f_0} &= \frac{1}{2}(\ket{e_0} + \ket{e_1} + \ket{e_2} + \ket{e_3})\label{eq:f-state}\\
\ket{f_1} &= \frac{1}{2}(\ket{e_0} - \ket{e_1} + \ket{e_2} - \ket{e_3})\notag\\
\ket{f_2} &= \frac{1}{2}(\ket{e_0} + \ket{e_1} - \ket{e_2} - \ket{e_3})\notag\\
\ket{f_3} &= \frac{1}{2}(\ket{e_0} - \ket{e_1} - \ket{e_2} + \ket{e_3})\notag
\end{align}
Naturally, unitarity of $U$ imposes certain conditions on these states, namely:
\begin{align}
&\braket{e_0|e_0} + \braket{e_1|e_1} = \braket{e_2|e_2} + \braket{e_3|e_3} = 1\label{eq:unitary-requirements}\\
&\braket{e_0|e_2} + \braket{e_1|e_3} = 0.\notag
\end{align}
These conditions will be important later.

Changing basis, we may write $\ket{e} = \alpha\ket{+} + \beta\ket{-}$, where:
\begin{align}
\alpha &= \frac{1}{\sqrt{2}}\left(\sqrt{\frac{1}{2}+b} + \sqrt{\frac{1}{2} - b}\right)\label{eq:alpha-beta}\\
\beta &= \frac{1}{\sqrt{2}}\left(\sqrt{\frac{1}{2}+b} - \sqrt{\frac{1}{2} - b}\right)\notag
\end{align}
Thus, by the linearity of $U$, we have:
\[
U\ket{e} = \ket{+,g_0} + \ket{-,g_1},
\]
where:
\begin{align}
\ket{g_0} &= \alpha\ket{f_0} + \beta\ket{f_2}\label{eq:g-state}\\
\ket{g_1} &= \alpha\ket{f_1} + \beta\ket{f_3}.\notag
\end{align}
Since it will be useful later, we will also write $\ket{g_1}$ in terms of the $\ket{e_i}$ states.  Let $X = \sqrt{1/2+b}$ and $Y=\sqrt{1/2-b}$.  Then:
\begin{align}
\ket{g_1} &= \alpha\ket{f_1} + \beta\ket{f_3} = \frac{1}{2}( [\alpha+\beta]\ket{e_0} - [\alpha+\beta]\ket{e_1} + [\alpha-\beta]\ket{e_2} - [\alpha-\beta]\ket{e_3})\notag\\
&=\frac{1}{2}(\sqrt{2}X\ket{e_0} - \sqrt{2}X\ket{e_1} + \sqrt{2}Y\ket{e_2} - \sqrt{2}Y\ket{e_3})\notag\\
&=\frac{1}{\sqrt{2}}(X\ket{e_0} - X\ket{e_1} + Y\ket{e_2} - Y\ket{e_3}).\label{eq:g1-termsof-e}
\end{align}
Observe that if $b=0$ we have $\ket{g_1} = \ket{f_1}$ as expected.

Thus, after $E$'s attack, the system evolves to:
\[
\rho = \frac{1}{2}\ketbra{0}_B \otimes P(\ket{+,g_0} + \ket{-,g_1}) + \frac{1}{2}\ketbra{1}_B \otimes P(\ket{0,e_0} + \ket{1,e_1}),
\]
where $P(z) = zz^*$ and $z^*$ represents the conjugate transpose of $z$.

$A$ will then receive the transit qubit and perform a measurement in either the $Z$ or $X$ basis.  She will accept only if she measures a $\ket{1}$ or a $\ket{-}$.  Thus, conditioning on her acceptance, the system becomes, disregarding the normalization term:
\begin{align*}
\sigma &= \ketbra{0}_A \otimes \left(\frac{1}{2}\ketbra{0}_B\otimes\frac{1}{2}P(\ket{g_0} - \ket{g_1}) + \frac{1}{2}\ketbra{1}_B \otimes\ketbra{e_1}\right)\\
&+ \ketbra{1}_A \otimes \left(\frac{1}{2}\ketbra{0}_B\otimes\ketbra{g_1} + \frac{1}{2}\ketbra{1}_B\otimes\frac{1}{2}P(\ket{e_0} - \ket{e_1}) \right).
\end{align*}

Rearranging terms, and now inserting the normalization term $N = tr\sigma$, yields the final system:
\begin{align}
\rho_{ABE} &= \frac{1}{N}\left[ \frac{1}{2}\ketbra{00}_{AB} \otimes \frac{1}{2}P(\ket{g_0} - \ket{g_1}) + \frac{1}{2}\ketbra{01}_{AB}\otimes\ketbra{e_1} \right.\label{eq:rho-ABE}\\
&\left.  +  \frac{1}{2}\ketbra{10}_{AB}\otimes\ketbra{g_1} + \frac{1}{2}\ketbra{11}_{AB} \otimes \frac{1}{2}P(\ket{e_0} - \ket{e_1})\right].\notag
\end{align}

Let $q_{i,j}$ be defined as follows:
\begin{align}
q_{0,0} &= \frac{1}{4}tr P(\ket{g_0} - \ket{g_1}) = \frac{1}{4}(1 - 2Re\braket{g_0|g_1})\label{eq:q-values}\\
q_{1,1} &= \frac{1}{4}tr P(\ket{e_0} - \ket{e_1}) = \frac{1}{4}(1 - 2Re\braket{e_0|e_1})\notag\\
q_{0,1} &= \frac{1}{2}\braket{e_1|e_1} = \frac{1}{2}Q_Z\notag\\
q_{1,0} &= \frac{1}{2}\braket{g_1|g_1} = \frac{1}{2}Q_e,\notag
\end{align}
where $Q_Z$ is the probability that a $\ket{0}$ flips to a $\ket{1}$ while $Q_e = |\braket{-|U|e}|^2$ is the probability that, if $B$ reflects and if $A$ measures in the $X$ basis, that she measures $\ket{-}$.  Notice that, when $b=0$, then $\ket{e} = \ket{+}$ and so $Q_e$ in that case is the error rate in the $X$ basis.  Of course the bias term affects this error rate and to stress that fact, we write $Q_e$ instead of $Q_X$.

With these definitions, clearly:
\begin{equation}
N = \sum_{i,j}q_{i,j}.
\end{equation}

Let $p_{i,j}$ be the probability that $A$ and $B$'s raw key bit is $i$ and $j$ respectively, conditioning on the event that $A$ and $B$ accept.  These values are obviously:
\begin{equation}\label{eq:p-values}
p_{i,j} = \frac{q_{i,j}}{N}
\end{equation}

\subsection{Bounding the von Neumann Entropy}

We will now compute a lower-bound on the key rate of this protocol.  Tracing out $A$'s system yields:
\begin{align*}
\rho_{BE} &= \frac{1}{N}\left[ \frac{1}{2}\ketbra{0}_B \otimes \left(\frac{1}{2}P(\ket{g_0} - \ket{g_1}) + \ketbra{g_1}\right)\right.\\
&\left. + \frac{1}{2}\ketbra{1}_B \otimes \left(\frac{1}{2}P(\ket{e_0} - \ket{e_1}) + \ketbra{e_1}\right)\right].
\end{align*}

Computing the von Neumann entropy of such a state is a difficult task.  To simplify the system, we will use a technique, inspired from \cite{QKD-keyrate-general} and also used in \cite{SQKD-Krawec-SecurityProof} to prove the security of Boyer et al.,'s \cite{SQKD-first} SQKD protocol.  That is, we will condition on a new random variable $C$.  Note that, due to the strong sub additivity of von Neumann entropy, for any tripartite system, it holds that $S(B|E) \ge S(B|EC)$.  Thus, if we condition on a new system $C$, we will derive a lower-bound on the key rate equation \ref{eq:keyrate}.  By careful choice of $C$ we will also simplify the entropy computation.

Our system $C$ will be two-dimensional, spanned by the orthonormal basis $\{\ket{C}, \ket{W}\}$.  Here, $\ketbra{C}$ will be the event that $A$ and $B$'s raw key bits match (i.e., they are Correct), while $\ketbra{W}$ will describe the event that $A$ and $B$'s raw key bits are Wrong.

Conditioning on this new system yields the mixed state:
\begin{align*}
\rho_{BEC} &= \frac{1}{N}\left[ \frac{1}{2}\ketbra{0}_B \otimes \left(\ketbra{C}\otimes \frac{1}{2}P(\ket{g_0} - \ket{g_1}) + \ketbra{W}\otimes\ketbra{g_1}\right)\right.\\
&\left. + \frac{1}{2}\ketbra{1}_B \otimes \left(\ketbra{C}\otimes\frac{1}{2}P(\ket{e_0} - \ket{e_1}) + \ketbra{W}\otimes\ketbra{e_1}\right)\right].
\end{align*}

Choosing a suitable basis, we may write $\rho_{BEC}$ as a diagonal matrix with diagonal elements of the form $p_{i,j}$.  Thus, we readily compute:
\begin{equation}\label{eq:S-BEC}
S(BEC) = H(p_{0,0}, p_{0,1}, p_{1,0}, p_{1,1}) = H\left(\left\{p_{i,j}\right\}_{i,j}\right),
\end{equation}
where $H(\cdot)$ is the Shannon entropy function.

We must now find an upper-bound on the quantity $S(EC)$ (thus providing us with a lower bound on $S(B|EC) = S(BEC) - S(EC)$).  Tracing out $B$ yields:
\begin{align}
\rho_{EC} &= \frac{1}{N}\left[ \ketbra{C} \otimes \frac{1}{4}( P(\ket{g_0} - \ket{g_1}) + P(\ket{e_0} - \ket{e_1}))\right.\label{eq:stateEC}\\
&\left.+\ketbra{W} \otimes \frac{1}{2}(\ketbra{g_1} + \ketbra{e_1})\right].\notag
\end{align}

Assume for now that $q_{0,1}$ and $q_{1,0}$ are both positive ($q_{0,0}$ and $q_{1,1}$ should both be positive, else there is too much noise and $A$ and $B$ should abort).  After some algebraic manipulation, we can rewrite the above density operator in the form:
\begin{equation}
\rho_{EC} = (p_{0,0} + p_{1,1}) \ketbra{C} \otimes \sigma_C + (p_{0,1} + p_{1,0})\ketbra{W}\otimes\sigma_W,
\end{equation}
where:
\begin{align*}
\sigma_C &= \frac{P(\ket{g_0} - \ket{g_1}) + P(\ket{e_0} - \ket{e_1})}{4(q_{0,0} + q_{1,1})}\\\\
\sigma_w &= \frac{\ketbra{g_1} + \ketbra{e_1}}{2(q_{0,1} + q_{1,0})}.
\end{align*}
Observe that both $\sigma_C$ and $\sigma_W$ are Hermitian positive semi-definite operators of unit trace.

For states of this form, it is easy to show (see \cite{SQKD-Krawec-SecurityProof} for a proof), that:
\begin{align*}
S(EC) &= h(p_{0,0} + p_{1,1}) + (p_{0,1} + p_{1,0})S(\sigma_W) + (p_{0,0} + p_{1,1})S(\sigma_C)\\
&\le h(p_{0,0} + p_{1,1}) + p_{0,1} + p_{1,0} + (p_{0,0} + p_{1,1})S(\sigma_C),
\end{align*}
where $h(x)$ is the binary entropy function (e.g., $h(x) = H(x, 1-x)$), and where the inequality follows from the fact that, since $\sigma_W$ is two-dimensional, $S(\sigma_W) \le 1$.

If the error rate is small, it is expected that $p_{0,1}$ and $p_{1,0}$ are also small (note that if both are zero, it is easy to show that $\ket{g_1} \equiv \ket{e_1} \equiv 0$; thus they never appear in Equation \ref{eq:stateEC} and the bound above applies even in this case).  However $p_{0,0}$ and $p_{1,1}$ should be large thus we must find an upper-bound on $S(\sigma_C)$.

Let $\ket{h_0} = \ket{g_0} - \ket{g_1}$ and $\ket{h_1} = \ket{e_0} - \ket{e_1}$.  Then $q_{0,0} = \braket{h_0|h_0}/4$ and $q_{1,1} = \braket{h_1|h_1}/4$ and:
\begin{align*}
\sigma_C = \frac{\ketbra{h_0} + \ketbra{h_1}}{\braket{h_0|h_0} + \braket{h_1|h_1}}.
\end{align*}

We may write, without loss of generality, $\ket{h_0} = x\ket{h}$ and $\ket{h_1} = y\ket{h} + z\ket{\zeta}$, where $x,y,z\in \mathbb{C}$, $\braket{h|h} = \braket{\zeta|\zeta} = 1$ and $\braket{h|\zeta} = 0$.  This implies:
\begin{align}
|x|^2 &= \braket{h_0|h_0} = 4q_{0,0}\\
|y|^2 + |z|^2 &= \braket{h_1|h_1} = 4q_{1,1}\\
x^*y &= \braket{h_0|h_1} \Longrightarrow |y|^2 = \frac{|\braket{h_0|h_1}|^2}{|x|^2}
\end{align}

In this $\{\ket{h}, \ket{\zeta}\}$ basis, we may write $\sigma_C$ as:
\[
\sigma_C = \frac{1}{|x|^2 + |y|^2 + |z|^2} \left(\begin{array}{ccc}
|x|^2 + |y|^2 &,& yz^*\\\\
y^*z &,& |z|^2\end{array}\right),
\]
the eigenvalues of which are:
\begin{align*}
\lambda_\pm &= \frac{1}{2} \pm \frac{\sqrt{(|x|^2 + |y|^2 - |z|^2)^2 + 4|y|^2|z|^2}}{2(|x|^2 + |y|^2 + |z|^2)}\\\\
&=\frac{1}{2} \pm \frac{\sqrt{(|x|^2 + 2|y|^2 - 4q_{1,1})^2 + 4|y|^2(4q_{1,1}-|y|^2)}}{2(|x|^2 + |y|^2 + |z|^2)},
\end{align*}
where, above, we used the fact that $|y|^2 + |z|^2 = 4q_{1,1}$.  Now, let $\Delta = 4q_{0,0} - 4q_{1,1}$ and, recalling that $|x|^2 = 4q_{0,0}$, we continue:

\begin{align*}
\lambda_\pm &= \frac{1}{2} \pm\frac{\sqrt{(\Delta + 2|y|^2)^2 + 16|y|^2q_{1,1} - 4|y|^4}}{2(|x|^2 + |y|^2 + |z|^2)}\\\\
&=\frac{1}{2} \pm\frac{\sqrt{\Delta^2 + 4|y|^2\Delta + 16|y|^2q_{1,1}}}{2(|x|^2 + |y|^2 + |z|^2)}\\\\
&=\frac{1}{2} \pm\frac{\sqrt{\Delta^2+4|y|^2(\Delta+4q_{1,1})}}{2(|x|^2 + |y|^2 + |z|^2)} = \frac{1}{2} \pm\frac{\sqrt{\Delta^2 + 4|y|^2(4q_{0,0})}}{8(q_{0,0} + q_{1,1})}.
\end{align*}
Recalling that $|y|^2 = |\braket{h_0|h_1}|^2/|x|^2 = |\braket{h_0|h_1}|^2/(4q_{0,0})$, yields:
\begin{equation}
\lambda_\pm = \frac{1}{2}\pm\frac{\sqrt{\Delta^2 + 4|\braket{h_0|h_1}|^2}}{8(q_{0,0} + q_{1,1})} = \frac{1}{2}\pm\frac{\sqrt{4(q_{0,0}-q_{1,1})^2 + |\braket{h_0|h_1}|^2}}{4(q_{0,0} + q_{1,1})}.
\end{equation}

Thus, we have:
\[
S(EC) \le h(p_{0,0} + p_{1,1}) + p_{0,1} + p_{1,0} + (p_{0,0} + p_{1,1})h\left(\lambda_+\right).
\]

This is a quantity which depends on the values $p_{i,j}$, which may be directly observed by $A$ and $B$, and also $|\braket{h_0|h_1}|^2$, a quantity which, though not directly observable, can be bounded as we will soon demonstrate.

Note that $\lambda_+ \ge 1/2$ and that $h(x)$ attains its maximum when $x = 1/2$.  Also observe that, as $|\braket{h_0|h_1}|^2\ge 0$ increases, $\lambda_+$ increases, thus decreasing $h(\lambda_+)$.  Therefore, if we find a lower bound $\mathcal{B}$ such that:
\[
|\braket{h_0|h_1}|^2 \ge \mathcal{B},
\]
and define:
\begin{equation}\label{eq:lambda}
\lambda = \frac{1}{2}+\frac{\sqrt{4(q_{0,0}-q_{1,1})^2 + \mathcal{B}}}{4(q_{0,0} + q_{1,1})},
\end{equation}
then $\frac{1}{2}\le\lambda \le \lambda_+$ which implies $h(\lambda) \ge h(\lambda_+)$ and so:
\[
S(EC) \le h(p_{0,0} + p_{1,1}) + p_{0,1} + p_{1,0} + (p_{0,0} + p_{1,1})h\left(\lambda\right).
\]

Thus, our key rate bound becomes:
\begin{equation}
r \ge H\left(\{p_{i,j}\}_{i,j}\right) - h(p_{0,0} + p_{1,1}) - p_{0,1} - p_{1,0} - (p_{0,0} + p_{1,1})h(\lambda) - H(B|A).
\end{equation}

\subsection{Final Key Rate Bound}

Computing $H(B|A)$ is trivial.  Indeed, $H(B|A) = H(BA) - H(A)$.  From Equation \ref{eq:rho-ABE}, it is easy to see that $H(BA) = H(\{p_{i,j}\}_{i,j})$.  $H(A)$ is simply $h(p_{0,0} + p_{0,1})$.  Thus our key rate bound becomes:
\begin{equation}\label{eq:final-keyrate}
r \ge h(p_{0,0} + p_{0,1}) - h(p_{0,0} + p_{1,1}) - p_{0,1} - p_{1,0} - (p_{0,0} + p_{1,1})h(\lambda).
\end{equation}

This is an equation which, with the exception of $\mathcal{B}$, depends only on parameters that $A$ and $B$ may estimate.  Determining a value for $\mathcal{B}$ may be done by considering certain other observable statistics as we demonstrate in the next section.

\subsection{Bounding $|\braket{h_0|h_1}|^2$ Using Certain Observable Statistics}
Our goal is to bound $|\braket{h_0|h_1}|^2 = |\braket{h_1|h_0}|^2$ (i.e., determine a value for $\mathcal{B}$ needed for evaluating Equation \ref{eq:lambda} which appears in our key rate bound in Equation \ref{eq:final-keyrate}).  While $|\braket{h_0|h_1}|^2$ cannot be observed directly, it may be bounded using only statistics that are directly observable by quantum $A$ and classical $B$.

Recall $\ket{h_0} = \ket{g_0} - \ket{g_1}$ and $\ket{h_1} = \ket{e_0} - \ket{e_1}$.  From Equations \ref{eq:f-state} and \ref{eq:g-state}, we have:
\begin{align*}
\ket{h_0} &= \ket{g_0} - \ket{g_1} = \alpha(\ket{f_0} - \ket{f_1}) + \beta(\ket{f_2} - \ket{f_3})\\
&= \alpha(\ket{e_1} + \ket{e_3}) + \beta(\ket{e_1} - \ket{e_3}),
\end{align*}
where $\alpha$ and $\beta$ are defined in Equation \ref{eq:alpha-beta}.

Let $X = \sqrt{1/2 + b}$ and $Y = \sqrt{1/2-b}$.  Then, writing $\alpha = \frac{1}{\sqrt{2}}(X+Y)$ and $\beta = \frac{1}{\sqrt{2}}(X-Y)$, we have:
\begin{align*}
\ket{h_0} = \frac{1}{\sqrt{2}}X(2\ket{e_1}) + \frac{1}{\sqrt{2}}Y(2\ket{e_3}) &= \sqrt{2}X\ket{e_1} + \sqrt{2}Y\ket{e_3}\\
&=\sqrt{1+2b}\ket{e_1} + \sqrt{1-2b}\ket{e_3}\\
&= \gamma\ket{e_1} + \delta\ket{e_3},
\end{align*}
where we have defined $\gamma = \sqrt{1+2b}$ and $\delta = \sqrt{1-2b}$.

In this new notation, we have:
\begin{align*}
\braket{h_1|h_0} &= \gamma\braket{e_0|e_1} - \gamma\braket{e_1|e_1} + \delta\braket{e_0|e_3} - \delta\braket{e_1|e_3}.
\end{align*}
(Note that the order of the $h$ states in the above - i.e., $\braket{h_1|h_0}$ instead of $\braket{h_0|h_1}$ - is not a typo.)

Clearly, Alice and Bob may estimate $\braket{e_1|e_1}$: it is simply the probability that, if $B$ sends a $\ket{0}$ (i.e., he measures and resends a $\ket{0}$), then $A$ measures $\ket{1}$.  Call this quantity $Q_Z$: it is the error rate in the $Z$ basis.  We will make the usual assumption in QKD security proofs that the error in the $Z$ basis is symmetrical in that $Q_Z = \braket{e_1|e_1} = \braket{e_2|e_2}$ (since $\braket{e_2|e_2}$ is also observable, this assumption could even be enforced); our analysis in the following section, however, may be carried out without this assumption.

Now consider:
\begin{equation}\label{eq:real-h}
Re\braket{h_1|h_0} = \gamma Re\braket{e_0|e_1} - \gamma Q_Z + \delta Re\braket{e_0|e_3} - \delta Re\braket{e_1|e_3},
\end{equation}
where $Re(z)$ denotes the real part of $z$.  If we can find a value $\eta \ge 0$ such that $Re\braket{h_1|h_0} \ge \eta \ge 0$, it will follow that:
\[
|\braket{h_0|h_1}|^2 = |\braket{h_1|h_0}|^2 = Re^2\braket{h_1|h_0} + Im^2\braket{h_1|h_0} \ge Re^2\braket{h_1|h_0} \ge \eta^2,
\]
where $Im(z)$ denotes the imaginary part of $z$.  Of course if $\eta$ is negative the above cannot be used; however, we will show, if the noise in the channel is low enough, $\eta$ is also large positive.  Thus, finding a positive lower bound for Equation \ref{eq:real-h} is our new goal.  To do so, $A$ and $B$ will use mismatched measurement results - a technique which will allow them to observe quantities such as $Re\braket{e_0|e_1}$ by using those iterations where $A$ measures in the ``wrong'' basis (e.g., they will use the probability that $A$ measures $\ket{+}$ if $B$ sent $\ket{0}$).  Using mismatched bases to estimate the quantum channel in quantum key distribution is not a new idea (see, for instance, \cite{QKD-mismatch1}).

Observe that, if $B$ sends $\ket{0}$ then, after $E$'s attack operator $U$, the state when the qubit arrives at $A$ is:
\[
U\ket{0} = \ket{0,e_0} + \ket{1,e_1} = \frac{1}{\sqrt{2}}\ket{+}(\ket{e_0} + \ket{e_1}) + \frac{1}{\sqrt{2}}\ket{-}(\ket{e_0} - \ket{e_1}).
\]
If $A$ measures in the $X$ basis, she observes $\ket{+}$ with probability:
\[
p_{0+} = \frac{1}{2}(1 + 2Re\braket{e_0|e_1}),
\]
thus providing her with an estimate of $Re\braket{e_0|e_1}$.  Note that, for notation, we will write $p_{i,j}$ to mean the probability that $A$ measures $j$ if $B$ sends $i$.  These are not to be confused with Equation \ref{eq:p-values}; and indeed, they cannot be since the $p_{\cdot, \cdot}$ values we consider in this section will always be ``mismatched'' (one of $i$ or $j$ will be a $Z$ state while the other will be a non $Z$ state).

To estimate $Re\braket{e_1|e_3}$, we can use the probability that $A$ measures $\ket{1}$ if $B$ reflects.  Recall, the state arriving at $B$'s lab is $\ket{e} = X\ket{0} + Y\ket{1}$ (where $X=\sqrt{1/2+b}$ and $Y = \sqrt{1/2-b}$).  After $E$'s attack operator in the return channel (assuming $B$ reflected), the state evolves to:
\[
U\ket{e} = X(\ket{0,e_0} + \ket{1,e_1}) + Y(\ket{0,e_2} + \ket{1,e_3}) = \ket{0}(X\ket{e_0} + Y\ket{e_2}) + \ket{1}(X\ket{e_1} + Y\ket{e_3}).
\]
Thus $p_{e1}$ - the probability that $A$ measures $\ket{1}$ if $B$ ``sends'' $\ket{e}$ (i.e., reflects) - is:
\begin{align*}
p_{e1} &= X^2\braket{e_1|e_1} + Y^2\braket{e_3|e_3} + 2XYRe\braket{e_1|e_3}\\
&= X^2Q_Z + (1-Q_Z)Y^2 + 2XYRe\braket{e_1|e_3}\\
&=\left(\frac{1}{2}+b\right)Q_Z + \left(\frac{1}{2}-b\right)(1-Q_Z) + 2\sqrt{\frac{1}{4}-b^2}Re\braket{e_1|e_3}\\
&=\frac{1}{2} - b(1-2Q_Z) + 2\sqrt{\frac{1}{4}-b^2}Re\braket{e_1|e_3}.
\end{align*}
(Note that if $b=0$ and $Q_Z=0$ - i.e, there is no error in either direction of the channel - then $p_{e1} = p_{+1} = 1/2$ as expected.)

Since $b$ and $Q_Z$ are parameters that may be estimated by $A$ and $B$, this value $p_{e1}$ provides them with an estimate of $Re\braket{e_1|e_3}$.

Finally, for $Re\braket{e_0|e_3}$, we consider the probability that $A$ measures $\ket{-}$ if $B$ reflects (note this should be small if the noise is small).  Using Equation \ref{eq:g1-termsof-e}, this value is found to be:
\begin{align}
p_{e-} &= Q_e = \braket{g_1|g_1} = \frac{1}{2}\left(X^2\braket{e_0|e_0} + X^2\braket{e_1|e_1} + Y^2\braket{e_2|e_2} + Y^2\braket{e_3|e_3}\right.\notag\\
&\left.- 2Re\left[X^2\braket{e_0|e_1} + XY\braket{e_0|e_3} + XY\braket{e_1|e_2} + Y^2\braket{e_2|e_3}\right]\right).\notag\\
\notag\\
&=\frac{1}{2} - Re\left(X^2\braket{e_0|e_1} + XY\braket{e_0|e_3} + XY\braket{e_1|e_2} + Y^2\braket{e_2|e_3}\right)\label{eq:pr-eminus}
\end{align}

Above, we used Equation \ref{eq:unitary-requirements} to make certain cancelations.

Note that $Re\braket{e_2|e_3}$ may be estimated in a similar manner as was $Re\braket{e_0|e_1}$, namely by using the value $p_{1+}$.  $Re\braket{e_1|e_2}$ may be bounded using the Cauchy-Schwarz inequality: $|\braket{e_1|e_2}| \le \sqrt{\braket{e_1|e_1}\braket{e_2|e_2}} = \sqrt{Q_Z^2} = Q_Z$.  Thus:
\[
Re\braket{e_1|e_2} \le \sqrt{Re^2\braket{e_1|e_2}} \le |\braket{e_1|e_2}| \le Q_Z.
\]

This provides $A$ and $B$ with a bound on the quantity $Re\braket{e_0|e_3}$.  We now have everything necessary to find a lower bound $\eta$ such that $Re\braket{h_1|h_0} \ge \eta$.  In particular, $\eta$ may be found using the (observable) parameters:
\begin{align}
p_{0+} &= \frac{1}{2}(1+2Re\braket{e_0|e_1})\label{eq:p0p-compute}\\
p_{1+} &= \frac{1}{2}(1+2Re\braket{e_2|e_3})\label{eq:p1p-compute}\\
p_{e1} &= \frac{1}{2} - b(1-2Q_Z) + 2\sqrt{\frac{1}{4}-b^2}Re\braket{e_1|e_3}\label{eq:pe1-compute}\\
p_{e-} &= \text{[from Equation \ref{eq:pr-eminus}]}.
\end{align}

Assuming $\eta \ge 0$, then we may set $\mathcal{B} = \eta^2$ and thus compute our key rate bound $r$ from Equation \ref{eq:final-keyrate}.  We will demonstrate how this may be done in the next section when we consider a specific attack scenario.  We will consider a specific attack scenario so as to provide us with numbers to put to these many variables so as to evaluate our key rate bound.  However, in practice, these values will be estimated from the observed statistics; our key rate bound derived in this paper applies even in the most general of scenarios.

\subsection{General Attacks}

In the previous section we considered collective attacks (technically restricted collective attacks, but these imply security against collective attacks for protocols of this type as proven in \cite{SQKD-Single-Security}).  However, since the protocol is permutation invariant \cite{QKD-renner-keyrate,QKD-renner-keyrate2}, the results from \cite{QKD-general-attack,QKD-general-attack2} apply and so to prove security against general attacks, it is sufficient to show security against collective attacks.  Furthermore, our key rate bound, in the asymptotic scenario, holds true even against arbitrary general attacks, thus giving us unconditional security.

\section{Evaluation}

Our work in the preceding section applies to the most general case.  Indeed, in practice, $A$ and $B$ will estimate the parameters mentioned, determine a value for $\eta$ and, assuming the noise is small enough, use that value to compute $r$.  In the interest of evaluating our work in this paper, however, we will demonstrate our key rate bound in a particular attack scenario.  Namely, we will assume that $E$'s reverse channel attack may be modeled by a depolarization channel with parameter $q$.  Such a channel acts on two-dimensional density operators $\rho$ as follows:
\[
\mathcal{E}_q(\rho) = (1-q)\rho + \frac{q}{2}I,
\]
where $I$ is the identity operator.  Note a depolarization channel is the typical scenario considered in the security proofs of B92 \cite{QKD-keyrate-general,QKD-B92-Improved}.


From our work in the previous section, to compute a lower bound on the key rate, we need to compute $Q_Z$, $p_{0+}$, $p_{1+}$, $p_{e1}$, and $p_{e-}$ given bias $b$.  We also need $q_{i,j}$ for $i,j \in \{0,1\}$ (see Equation \ref{eq:q-values}) which will give us $p_{i,j}$ and $N$.

The first three values are easily computed.  Indeed, if $B$ sends $\ket{i}$ for $i \in \{0,1\}$, then the state arriving at $A$'s lab is:
\begin{equation}\label{eq:dep-send}
\mathcal{E}_q(\ketbra{i}) = (1-q)\ketbra{i} + \frac{q}{2}(\ketbra{i} + \ketbra{1-i}),
\end{equation}
and so $Q_Z = q/2$ and:
\[
p_{0+} = p_{1+} = \frac{1-q}{2} + \frac{q}{4} + \frac{q}{4} = \frac{1}{2}.
\]
These last two imply that $Re\braket{e_0|e_1} = Re\braket{e_2|e_3} = 0$ (from Equations \ref{eq:p0p-compute} and \ref{eq:p1p-compute}).

Next, we compute $p_{e1}$.  Let $\ket{e} = \sqrt{1/2+b}\ket{0} + \sqrt{1/2-b}\ket{1}$ as before and let $\ket{\bar{e}}$ be a normalized state such that $\braket{\bar{e}|e} = 0$.  Under such conditions it is trivial to show that $\braket{0|\bar{e}} = \sqrt{1/2-b}$ and $\braket{1|\bar{e}} = \sqrt{1/2+b}$.  Then:
\begin{equation}\label{eq:dep-reflect}
\mathcal{E}_q(\ketbra{e}) = (1-q)\ketbra{e} + \frac{q}{2}(\ketbra{e} + \ketbra{\bar{e}}).
\end{equation}
From this it is clear that:
\[
p_{e1} = (1-q)\left(\frac{1}{2}-b\right) + \frac{q}{2}\left(\frac{1}{2}-b + \frac{1}{2}+b\right) = \frac{1}{2}-b(1-q) = \frac{1}{2}-b(1-2Q_Z).
\]
So long as $b \in (-1/2, 1/2)$ (i.e., $b \ne \pm 1/2$), this implies, from Equation \ref{eq:pe1-compute}, that $Re\braket{e_1|e_3} = 0$.

Next, we consider $p_{e-} = Q_e$ which is the probability that $A$ measures $\ket{-}$ if $B$ reflects.  In the absence of noise, this should be $0$.  From Equation \ref{eq:pr-eminus} and the above work, we have:
\begin{align*}
p_{e-} = Q_e = \braket{g_1|g_1} &= \frac{1}{2} - XYRe(\braket{e_0|e_3} + \braket{e_1|e_2})
\end{align*}
which implies:
\begin{align*}
Re\braket{e_0|e_3} = \frac{1-2Q_e}{2XY} - Re\braket{e_1|e_2} = \frac{1-2Q_e}{2\sqrt{\frac{1}{4}-b^2}} - Re\braket{e_1|e_2}.
\end{align*}
By applying the Cauchy-Schwarz inequality, as discussed in the previous section, we have $Re\braket{e_1|e_2} \le Q_Z$ and so:
\[
Re\braket{e_0|e_3} \ge \frac{1-2Q_e}{2\sqrt{\frac{1}{4}-b^2}} - Q_Z.
\]
(Note that if there is no noise and no bias, then $Re\braket{e_0|e_3} = 1$.)

Thus, we conclude that:
\begin{align*}
Re\braket{h_1|h_0} &= \gamma Re\braket{e_0|e_1} - \gamma Q_Z + \delta Re\braket{e_0|e_3} - \delta Re\braket{e_1|e_3}\\
&= \delta Re\braket{e_0|e_3} - \gamma Q_Z\\\\
&\ge \sqrt{1-2b}\left(\frac{1-2Q_e}{2\sqrt{\frac{1}{4}-b^2}} - Q_Z\right) - Q_Z\sqrt{1+2b} = \eta.
\end{align*}
So long as $\eta \ge 0$, this may be used to bound $|\braket{h_0|h_1}|^2 \ge \eta^2$.

We now need to compute $q_{i,j}$ and $Q_e$ in terms of $q$.  It is clear that $q_{0,0}$ is the probability that $A$ measures $\ket{1}$ and $B$ chooses to reflect (conditioning on the event that $B$ does not later discard this reflection iteration).  From Equation \ref{eq:dep-reflect}, this quantity is:
\[
q_{0,0} = \frac{1}{2} \left( (1-q)\left(\frac{1}{2}-b\right) + \frac{q}{2}\right) = \frac{1}{2}\left(\frac{1}{2}-b(1-q)\right).
\]

Similarly, $q_{1,1}$ is the probability that $A$ measures $\ket{-}$ and $B$ chooses to measure and resend, conditioning on the event he measures (and thus sends) $\ket{0}$.  From Equation \ref{eq:dep-send}, this is found to be:
\[
q_{1,1} = \frac{1}{2}\left(\frac{1-q}{2} + \frac{q}{2}\right) = \frac{1}{4}.
\]

The quantity $q_{0,1}$ is the probability that $A$ measures $\ket{1}$ and $B$ chooses to measure and resend, conditioning on the event he measures and sends $\ket{0}$.  So:
\[
q_{0,1} = \frac{1}{2}Q_Z = \frac{q}{4}.
\]

Finally, $q_{1,0}$ is the probability that $A$ measures $\ket{-}$ and $B$ chooses to reflect (again conditioning on the event he does not discard this reflection iteration).  This is found to be:
\[
q_{1,0} = \frac{1}{2}\underbrace{\left((1-q)\left(\frac{1}{2} - \sqrt{\frac{1}{4}-b^2}\right) + \frac{q}{2}\right)}_{Q_e} = \frac{1}{2}Q_e.
\]

This allows us to evaluate our key rate bound $r$.  The reader will observe that, though for this specific evaluation, we assumed a specific scenario of a depolarization channel in the reverse direction, all parameters used to make the necessary bounds are directly observable by $A$ and $B$; indeed, they may even enforce this particular scenario (as was the assumption in \cite{QKD-keyrate-general} when proving the security of B92 and BB84).

\begin{figure}
  \centering
  \includegraphics{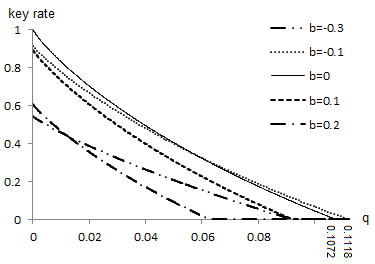}
\caption{A graph of our lower bound on the key rate of this SQKD protocol as a function of the depolarization channel parameter $q$ (note that, in this case, $Q_Z = q/2$) for various levels of bias $b$.  Note that, when $b=0$, the key rate remains positive for all $q \le .1072$ (i.e, $Q_Z \le 5.36\%$).  Note also that there are certain levels of bias which are beneficial to $A$ and $B$ (namely negative biases up to a certain amount).  In particular, if $b=-0.1$, the protocol's key rate remains positive for all $q \le .1118$ (i.e., $Q_Z \le 5.59\%$).  See the text for a discussion of this.}\label{fig:keyrate1}
\end{figure}

A graph of the lower-bound $r$ as a function of $q$ for various levels of bias $b$ is shown in Figure \ref{fig:keyrate1}.  When $b = 0$, our key rate remains positive for all $q \le .1072$ (or $Q_Z \le 5.36\%$).  This is comparable to $B92$ which, given an optimal choice of states, can tolerate up to $6.5\%$ \cite{QKD-B92-Improved} (note that, if we were to alter the protocol to use not $X$ basis states, but an optimally chosen basis, as is done with B92 \cite{QKD-keyrate-general}, we might be able to improve our key rate bound; we chose not to alter the protocol, however, in this manner for this paper and security analysis).  It is also comparable to the three state BB84 which can withstand up to $5.1\%$ error \cite{QKD-BB84-three-state-v2}.

\begin{figure}
  \centering
  \includegraphics{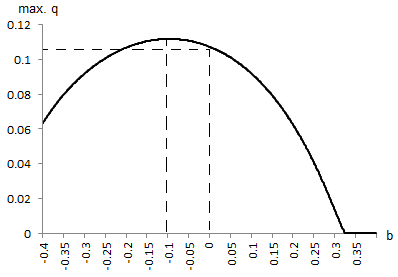}
\caption{Showing maximum $q$ for which our key rate bound remains positive as $b$ varies (said differently, the smallest $q$ at which our bound becomes zero).  When $b=0$, the rate remains positive for all $q \le .1072$.  When $b = -0.1$, it remains positive for all $q \le .1118$ (i.e., $Q_Z \le 5.59\%$).  When $b \ge .325$, our key rate bound is zero or negative for all $q$.  See the text for a discussion of this.}\label{fig:keyrate2}
\end{figure}

Perhaps most surprisingly is the observation that for small negative bias values (e.g., $b = -0.1$), while the key rate is worse for small values of $q$, it actually remains positive slightly longer, for greater values of $q$.  This is shown more clearly in Figure \ref{fig:keyrate2} which shows a graph of the function:
\[
\tau_Q(b) = \inf \{q \text{ } | \text{ } r \le 0 \text{ given bias $b$}\},
\]
as $b$ varies (i.e., this figure depicts the maximally tolerated error rate, according to our lower bound, as a function of the forward channel attack parameter $b$).

\begin{figure}
  \centering
  \includegraphics{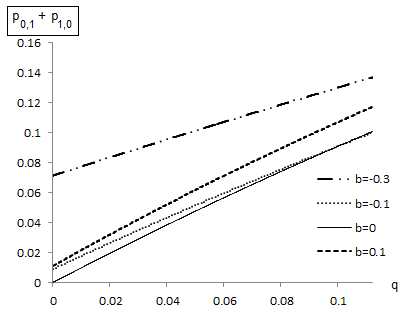}
\caption{Showing a graph of the value $p_{0,1} + p_{1,0}$, which is the error rate in $A$ and $B$'s raw key, as $q$ increases for various levels of $b$.  Note that $b=0$ produces the smallest such error except for large $q$ when it actually produces more error than when $b=-0.1$ (see also Figure \ref{fig:QBER1zoom}).  Positive values of $b$ generally produce the most error.}\label{fig:QBER1}
\end{figure}

\begin{figure}
  \centering
  \includegraphics{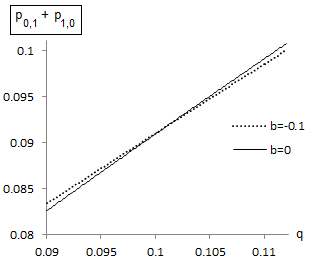}
\caption{A close up of the quantity $p_{0,1} + p_{1,0}$ (the error rate in $A$ and $B$'s raw key) for larger $q$.  Here it is more clear that there is a threshold after which $b=-0.1$ actually produces less error than $b=0$ (though the difference is slight, it does contribute to the difference in key rate bounds for the two bias values).  This contributes to the fact that when the bias is small negative the protocol can suffer a higher noise rate $q$.}\label{fig:QBER1zoom}
\end{figure}

\begin{figure}
  \centering
  \includegraphics{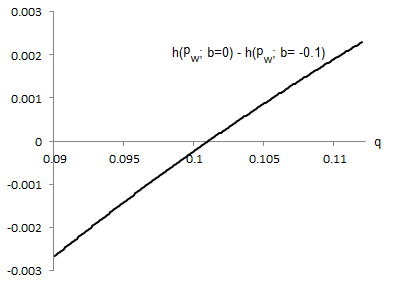}
\caption{Here we write $h(p_W ; b=0)$ to be the entropy $h(p_{0,1} + p_{1,0}) = h(p_{0,0} + p_{1,1})$ when bias $b=0$ is used.  Similarly we define $h(p_W;b=-0.1)$ to be the same computation but when bias $b=-0.1$ is used.  Observe that for larger $q$, the entropy when $b=0$ is larger than when $b=-0.1$.  Since this entropy term is subtracted from the key rate, this fact also contributes to the higher tolerated error rate of $b=-0.1$ (when $b=0$ we are subtracting a larger amount from $r$ for these higher values of $q$).}\label{fig:deltaH}
\end{figure}

\begin{figure}
  \centering
  \includegraphics{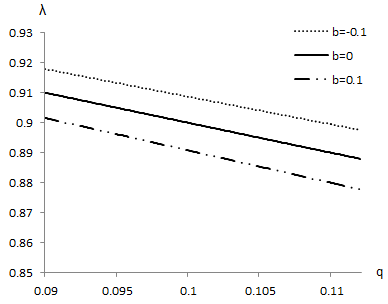}
\caption{A graph of the value $\lambda$ (Equation \ref{eq:lambda}) for various levels of bias $b$ and noise $q$.  Note that when $b=-0.1$, $\lambda$ is larger, thus $S(EC)$ (which depends on $h(\lambda)$) is smaller, thus increasing the key rate slightly.}\label{fig:lambda}
\end{figure}

To better understand this, we consider the effects of the bias on certain parameters.  First, we observe that when $b = -0.1$, the error rate of $A$ and $B$'s raw key (that is, the value $p_{0,1} + p_{1,0}$) is actually smaller for higher values of $q$ than the same value of $q$ given bias $b=0$.  This is depicted in Figures \ref{fig:QBER1} and \ref{fig:QBER1zoom}.  This is a small difference, however, since the quantity $h(p_{0,1}+p_{1,0}) = h(p_{0,0}+p_{1,1})$ appears in our key rate expression of Equation \ref{eq:final-keyrate}, this difference does contribute to the greater tolerated error rate for $b=-0.1$.  The difference between $h(p_{0,0}+p_{1,1})$ for $b = 0$ and $h(p_{0,0}+p_{1,1})$ when $b=-0.1$ is shown in Figure \ref{fig:deltaH}.  Finally, we observe that the value $\lambda$, in this particular scenario of a depolarization channel, is greater for $b = -0.1$ than it is for $b=0$ for large $q$.  The value of $\lambda$ is shown in Figure \ref{fig:lambda}.

Furthermore, consider B92, where $A$ and $B$ do not send $\ket{0}$ and $\ket{+}$ but instead choose two nonorthogonal states; different choices lead to better or worse performance.  Similarly here, we may be able to improve the key rate of our protocol even further by having $A$ not measure in the $X$ basis, but instead measure in an optimally chosen, non $Z$ basis  ($A$ would also send, not $\ket{+}$ but instead one of the states in this optimally chosen basis).  This could lead to interesting future work.  Note that our security analysis would apply even in this case, though of course the algebra would only be slightly different.  Of course, the parameter $b$ is in $E$'s control; clearly she would choose a positive value for this quantity, decreasing the key rate while increasing her advantage.  It is an open question, consider the fact that E controls the forward channel, as to how, if at all, an optimal choice of basis would alter the key rate expression (as it does for B92).

\section{Conclusion}
In this paper, we have provided a proof of unconditional security for a particular semi-quantum key distribution protocol.  We derived a lower bound on the key rate of this protocol in the asymptotic scenario - our lower bound is a function only of parameters that may be estimated by $A$ and $B$.  Finally, we evaluated this bound in a particular example: the depolarization channel.  This evaluation showed that this SQKD protocol compares favorably to certain fully quantum protocols.  That is to say, this SQKD protocol's maximally tolerated error rate of $5.36\%$ (the error rate threshold which determines when $A$ and $B$ must abort) is comparable to not only B92, but also the three state BB84 \cite{QKD-BB84-three-state} as we mentioned in the previous section.  This is an observation we've made concerning other semi-quantum cryptographic protocols \cite{SQKD-MultiUser,SQKD-Krawec-SecurityProof} yielding further evidence that, at least in this perfect qubit scenario, semi-quantum protocols are comparable to fully quantum ones.  Naturally, leaving the perfect qubit scenario is a challenge - as it is with any fully quantum protocol - especially those quantum protocols utilizing a two-way quantum channel.  This problem we leave as important future work, however we believe the proof techniques we develop here and in our prior work can greatly aid in this effort.  We also think, as mentioned before, our proof techniques can find application outside the realm of semi-quantum protocols to the proof of security for fully quantum protocols relying on two-way quantum channels.


\begin{thebibliography}{10}

\bibitem{QKD-BB84}
Charles~H Bennett and Gilles Brassard.
\newblock Quantum cryptography: Public key distribution and coin tossing.
\newblock In {\em Proceedings of IEEE International Conference on Computers,
  Systems and Signal Processing}, volume 175. New York, 1984.

\bibitem{QKD-B92}
Charles~H. Bennett.
\newblock Quantum cryptography using any two nonorthogonal states.
\newblock {\em Phys. Rev. Lett.}, 68:3121--3124, May 1992.

\bibitem{QKD-SARG04}
Antonio Acin, Nicolas Gisin, and Valerio Scarani.
\newblock Coherent-pulse implementations of quantum cryptography protocols
  resistant to photon-number-splitting attacks.
\newblock {\em Phys. Rev. A}, 69:012309, Jan 2004.

\bibitem{QKD-BB84-three-state}
Chi-Hang~Fred Fung and Hoi-Kwong Lo.
\newblock Security proof of a three-state quantum-key-distribution protocol
  without rotational symmetry.
\newblock {\em Phys. Rev. A}, 74:042342, Oct 2006.

\bibitem{QKD-survey}
Valerio Scarani, Helle Bechmann-Pasquinucci, Nicolas~J. Cerf, Miloslav
  Du\ifmmode~\check{s}\else \v{s}\fi{}ek, Norbert L\"utkenhaus, and Momtchil
  Peev.
\newblock The security of practical quantum key distribution.
\newblock {\em Rev. Mod. Phys.}, 81:1301--1350, Sep 2009.

\bibitem{SQKD-first}
Michel Boyer, D.~Kenigsberg, and T.~Mor.
\newblock Quantum key distribution with classical bob.
\newblock In {\em Quantum, Nano, and Micro Technologies, 2007. ICQNM '07. First
  International Conference on}, pages 10--10, 2007.

\bibitem{SQKD-second}
Michel Boyer, Ran Gelles, Dan Kenigsberg, and Tal Mor.
\newblock Semiquantum key distribution.
\newblock {\em Phys. Rev. A}, 79:032341, Mar 2009.

\bibitem{SQKD-3}
Wang Jian, Zhang Sheng, Zhang Quan, and Tang Chao-Jing.
\newblock Semiquantum key distribution using entangled states.
\newblock {\em Chinese Physics Letters}, 28(10):100301, 2011.

\bibitem{SQKD-4}
Kun-Fei Yu, Chun-Wei Yang, Ci-Hong Liao, and Tzonelih Hwang.
\newblock Authenticated semi-quantum key distribution protocol using bell
  states.
\newblock {\em Quantum Information Processing}, pages 1--9, 2014.

\bibitem{SQKD-lessthan4}
Xiangfu Zou, Daowen Qiu, Lvzhou Li, Lihua Wu, and Lvjun Li.
\newblock Semiquantum-key distribution using less than four quantum states.
\newblock {\em Phys. Rev. A}, 79:052312, May 2009.

\bibitem{SQKD-multi1}
Zhang Xian-Zhou, Gong Wei-Gui, Tan Yong-Gang, Ren Zhen-Zhong, and Guo
  Xiao-Tian.
\newblock Quantum key distribution series network protocol with m-classical
  bobs.
\newblock {\em Chinese Physics B}, 18(6):2143, 2009.

\bibitem{SQKD-Single-Security}
W.O. Krawec.
\newblock Restricted attacks on semi-quantum key distribution protocols.
\newblock {\em Quantum Information Processing}, 13(11):2417--2436, 2014.

\bibitem{SQKD-MultiUser}
Walter~O Krawec.
\newblock Mediated semiquantum key distribution.
\newblock {\em Physical Review A}, 91(3):032323, 2015.

\bibitem{SQKD-information}
Takayuki Miyadera.
\newblock Relation between information and disturbance in quantum key
  distribution protocol with classical alice.
\newblock {\em Int. J. of Quantum Information}, 9, 2011.

\bibitem{SQKD-cl-A}
Hua Lu and Qing-Yu Cai.
\newblock Quantum key distribution with classical alice.
\newblock {\em International Journal of Quantum Information}, 6(06):1195--1202,
  2008.

\bibitem{SQKD-Krawec-SecurityProof}
Walter~O Krawec.
\newblock Security proof of a semi-quantum key distribution protocol.
\newblock {\em to appear: IEEE ISIT 2015; arXiv preprint arXiv:1412.0282},
  2015.

\bibitem{SQKD-Krawec-dissertation}
Walter~O Krawec.
\newblock {\em Semi-Quantum Key Distribution}.
\newblock PhD thesis, Stevens Institute of Technology, May 2015.

\bibitem{SQKD-photon-tag}
Yong-gang Tan, Hua Lu, and Qing-yu Cai.
\newblock Comment on Òquantum key distribution with classical bobÓ.
\newblock {\em Phys. Rev. Lett.}, 102:098901, Mar 2009.

\bibitem{SQKD-photon-tag-comment}
Michel Boyer, Dan Kenigsberg, and Tal Mor.
\newblock Boyer, kenigsberg, and mor reply:.
\newblock {\em Phys. Rev. Lett.}, 102:098902, Mar 2009.

\bibitem{QKD-renner-keyrate}
Renato Renner, Nicolas Gisin, and Barbara Kraus.
\newblock Information-theoretic security proof for quantum-key-distribution
  protocols.
\newblock {\em Phys. Rev. A}, 72:012332, Jul 2005.

\bibitem{QKD-renner-keyrate2}
B.~Kraus, N.~Gisin, and R.~Renner.
\newblock Lower and upper bounds on the secret-key rate for quantum key
  distribution protocols using one-way classical communication.
\newblock {\em Phys. Rev. Lett.}, 95:080501, Aug 2005.

\bibitem{QKD-Winter-Keyrate}
Igor Devetak and Andreas Winter.
\newblock Distillation of secret key and entanglement from quantum states.
\newblock {\em Proceedings of the Royal Society A: Mathematical, Physical and
  Engineering Science}, 461(2053):207--235, 2005.

\bibitem{QKD-keyrate-general}
Matthias Christandl, Renato Renner, and Artur Ekert.
\newblock A generic security proof for quantum key distribution.
\newblock {\em arXiv preprint quant-ph/0402131}, 2004.

\bibitem{QKD-mismatch1}
Shun Watanabe, Ryutaroh Matsumoto, and Tomohiko Uyematsu.
\newblock Tomography increases key rates of quantum-key-distribution protocols.
\newblock {\em Physical Review A}, 78(4):042316, 2008.

\bibitem{QKD-general-attack}
Matthias Christandl, Robert Konig, and Renato Renner.
\newblock Postselection technique for quantum channels with applications to
  quantum cryptography.
\newblock {\em Phys. Rev. Lett.}, 102:020504, Jan 2009.

\bibitem{QKD-general-attack2}
Renato Renner.
\newblock Symmetry of large physical systems implies independence of
  subsystems.
\newblock {\em Nature Physics}, 3(9):645--649, 2007.

\bibitem{QKD-B92-Improved}
Ryutaroh Matsumoto.
\newblock Improved asymptotic key rate of the b92 protocol.
\newblock In {\em Information Theory Proceedings (ISIT), 2013 IEEE
  International Symposium on}, pages 351--353. IEEE, 2013.

\bibitem{QKD-BB84-three-state-v2}
Cyril Branciard, Nicolas Gisin, Norbert Lutkenhaus, and Valerio Scarani.
\newblock Zero-error attacks and detection statistics in the coherent one-way
  protocol for quantum cryptography.
\newblock {\em Quantum Information \& Computation}, 7(7):639--664, 2007.

\end{thebibliography}

\end{document}